\documentclass{article}

\usepackage{arxiv}

\usepackage[utf8]{inputenc} % allow utf-8 input
\usepackage[T1]{fontenc}    % use 8-bit T1 fonts
\usepackage{hyperref}       % hyperlinks
\usepackage{url}            % simple URL typesetting
\usepackage{booktabs}       % professional-quality tables
\usepackage{amsfonts}       % blackboard math symbols
\usepackage{nicefrac}       % compact symbols for 1/2, etc.
\usepackage{microtype}      % microtypography
\usepackage{lipsum}		% Can be removed after putting your text content
\usepackage{graphicx}
\usepackage{natbib}
\usepackage{doi}
\usepackage{amsmath}
\usepackage{wrapfig}
\usepackage{float}
\usepackage{afterpage}
\usepackage{comment}

\usepackage{tikz}
\usetikzlibrary{shapes.geometric, arrows.meta, positioning, trees, fit, backgrounds}

\usepackage{listings}
\usepackage{xcolor}
\usepackage{enumitem}

\definecolor{myblue}{HTML}{24abff} % #24abff
\definecolor{myyellow}{HTML}{ffa929} % #ffa929
\definecolor{myblued}{HTML}{005b94} % #005b94
\definecolor{myyellowd}{HTML}{945900} % #945900

\definecolor{green1}{RGB}{0, 139, 0}     % dark green
\definecolor{blue1}{RGB}{0, 0, 200}     % dark blue

\tikzstyle{fg} = [rectangle, rounded corners, minimum width=2cm, minimum height=1cm,text centered, draw=black, fill=myblue!50]
\tikzstyle{obs} = [rectangle, rounded corners, minimum width=2cm, minimum height=1cm,text centered, draw=black,, fill=myblue!50]
\tikzstyle{ana} = [rectangle, rounded corners, minimum width=2cm, minimum height=1cm,text centered, draw=black, fill=myblue!50]
\tikzstyle{model} = [rectangle, minimum width=2cm, minimum height=1cm, text centered, draw=black, fill=myyellow!50]
\tikzstyle{diff} = [rectangle, minimum width=2cm, minimum height=1cm, text centered, draw=black, fill=myyellow!50]
\tikzstyle{arrow} = [thick,->,>=stealth]

\hypersetup{colorlinks=true,
linkcolor=blue,
citecolor=blue,
plainpages=false}

\newcommand{\bx}{\mathbf{x}}
\newcommand{\hbx}{\hat{\bx}}

\title{AI-based data assimilation: \\Learning the functional of analysis estimation}
\author{ \href{https://orcid.org/0000-0002-2010-7767}{\includegraphics[scale=0.06]{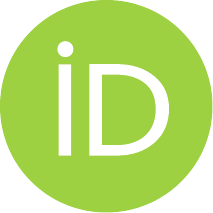}\hspace{1mm}Jan D. Keller}\thanks{\url{www.dwd.de}} \\
	Research and Development\\
	Deutscher Wetterdienst\\
	Offenbach, Germany \\
	\texttt{jan.keller@dwd.de} \\
	\And
	\href{https://orcid.org/0000-0001-6794-2500}
 {\includegraphics[scale=0.06]{orcid.pdf}\hspace{1mm}Roland Potthast} \\
	Research and Development\\
	Deutscher Wetterdienst\\
	Offenbach, Germany \\
	\texttt{roland.potthast@dwd.de} \\
}

\date{}

\renewcommand{\shorttitle}{Keller and Potthast: AI-based data assimilation}

\hypersetup{
pdftitle={AI-based data assimilation},
pdfsubject={},
pdfauthor={Jan D. Keller, Roland Potthast},
pdfkeywords={AI, data assimilation},
}

\begin{document}
\maketitle

%\fontsize{11pt}{14pt}\selectfont % arxive package rewrites default font size

\begin{abstract}
The integration of observational data into numerical models, known as data assimilation, is fundamental for making Numerical Weather Prediction (NWP) possible, with breathtaking success over the past 60 years (\cite{Bauer2015}). Traditional data assimilation methods, such as variational techniques and ensemble Kalman filters, are basic pillars of current NWP by incorporating diverse observational data. However, the emergence of artificial intelligence (AI) presents new opportunities for further improvements. AI-based approaches can emulate the complex computations of traditional NWP models at a reduced computational cost, offering the potential to speed up and improve analyses and forecasts dramatically (e.g.\ \cite{Pathak2022, Bi2023, Lam2023, ECMWF_AIFS_2024}). AI itself plays a growing role in optimization (e.g.\ \cite{fan2024artificial}), which offers new possibilities also beyond model emulation. 

In this paper, we introduce a novel AI-based variational data assimilation approach (AI-Var) designed to replace classical methods of data assimilation by leveraging deep learning techniques. Unlike previous hybrid approaches, our method integrates the data assimilation process directly into a neural network, utilizing the variational data assimilation framework. This innovative AI-based system, termed AI-Var, employs a neural network trained to minimize the variational cost function, enabling it to perform data assimilation without relying on pre-existing analysis datasets.

We present a proof-of-concept implementation of this approach, demonstrating its feasibility through a series of idealized and real-world test cases. Our results indicate that the AI-Var system can efficiently assimilate observations and produce accurate initial conditions for NWP, highlighting its potential to carry out the data assimilation process in weather forecasting. This advancement paves the way for fully data-driven NWP systems, offering a significant leap forward in computational efficiency and flexibility.
\end{abstract}

% keywords can be removed
\keywords{AI \and neural networks \and data assimilation \and variational methods \and reanalysis \and self-supervised learning \and optimization}

%===========================================================================
%
%===========================================================================
\section{Introduction}

In the dynamic and intricate field of Numerical Weather Prediction (NWP), the integration of observational data into numerical models stands as a cornerstone for making high forecast accuracy and reliability possible. This process, known as data assimilation, melds observations with model outputs to furnish refined analyses and initial conditions, thereby making the predictive prowess of NWP systems with high quality feasible. Traditional methods such as variational data assimilation techniques, the ensemble Kalman filter and its variants and recent developments on particle filters are the basis of current systems in this domain. There is a breathtaking improvement of forecast accuracy through algorithmical improvements and by adeptly incorporating a wide range of diverse observational data, e.g.\ \citep{Houtekamer1998,Houtekamer2001,Houtekamer2005,Anderson2001,lorenc_etal_2000,Lorenc2003, Kalnay2003, Hunt2007, Evensen2009, Anderson2012, Leeuwen2015, Reich2015, Nakamura2015, Bishop2016, Vetra2018, Leeuwen2019, Potthast2019}.

Current developments in artificial intelligence (AI)-based approaches have the potential to considerably enhance NWP \citep[e.g.,][]{Lam2023, Bi2023}. 
These sophisticated tools, leveraging deep learning techniques, are designed to replicate the complex physics-based calculations of traditional NWP models but at a significantly reduced computational cost for carrying out the forecasting step (called inference in the framework of machine learning). These NWP emulators unlock a potential not previously thought possible in terms of speeding up the forecasts generated by weather forecasting models. Weather predictions with lead times of a week or more normally taking hours to be performed can be calculated in seconds by AI emulators. This advancement not only expedites the NWP cycle, but also ensures that meteorologists can deliver timely and reliable weather predictions. On the other hand, classical data assimilation methods can thus become a bottleneck in the weather forecasting chain - taking up a much larger percentage of time needed to perform a NWP cycle. As the volume of meteorological data continues to grow, the gap in computing time needed between AI-based NWP models and the data assimilation step may even further increase.

Therefore, the advent of AI in NWP should also indicate a new era in data assimilation strategies, fostering the development of systems that intertwine well-tested traditional methodologies with the cutting-edge potential of machine learning. These innovative models aim to transcend the limitations of conventional techniques, offering a leap forward in refining initial conditions for NWP through AI-augmented assimilation processes \citep{Gottwald_2021, Dong2023, Arcucci2021, Bonavita2021}. One example is the proposition of merging machine learning with 4D-Var, suggesting a seamless integration of AI into the fabric of established data assimilation frameworks integrates deep learning methods into the data assimilation process. Already now, continuous learning of  uncertain components of the forecasting model based on observations have been introduced into operational state-of-the-art forecasting systems such as the ICON model of DWD \citep{Zaengl2023}, leading to much improved forecasting for surface variables. An important part of data assimilation systems is the mapping between model fields and the observations, known as {\em observation operators}. Using AI for improving or constructing observation operators has been a topic of research for many years, leading e.g. to an improved simulation of satellite radiances within RTTOV \citep{Scheck2021,Baur2023}, or for assimilating cloud images \citep{Reinhardt2023}. 

\begin{figure}[ht]
\begin{center}
\begin{tikzpicture}[level distance=1.5cm, sibling distance=10em,
  every node/.style = {shape=rectangle, rounded corners,
    draw, align=center,
    font=\sffamily, minimum height=1cm},  
    edge from parent/.style={draw, -latex, thick}]
  \node[fill=myblue!60] {DA Methods}
    child { node[fill=myblue!60,text width=3cm] (variational) {Variational Methods} 
      % Begin scope with a different sibling distance
      child [sibling distance=6em, text width = 1.5cm] { node[fill=myblue!60] (3dvar) {3D-Var} }
      child [sibling distance=6em, text width = 1.5cm] { node[fill=myblue!20] {4D-Var} } 
      child [sibling distance=6em, text width = 1.5cm] { node[fill=myblue!20] (envar) {EnVar} } 
    }
    child { node[fill=myblue!20,text width=3cm] {Ensemble Kalman Filters} }
    child { node[fill=myblue!20,text width=3cm] {Particle Filters} };

  % Dashed box around the specific nodes
  \begin{pgfonlayer}{background}
    \node[fit=(variational) (envar) (3dvar), draw, dashed, inner sep=0.3cm, fill=myblue!10] (box) {};
      \node[left=0.1cm of box, fill=myblue!60, draw=black, rounded corners, inner sep=0.2cm] (aivar) {AI-Var};

  \end{pgfonlayer}
  
\end{tikzpicture}
\end{center}
\caption{\label{fig:da_methods}
We will approach the task to develop an AI-based data assimilation method based on the 3D-Var functional.}
\end{figure}
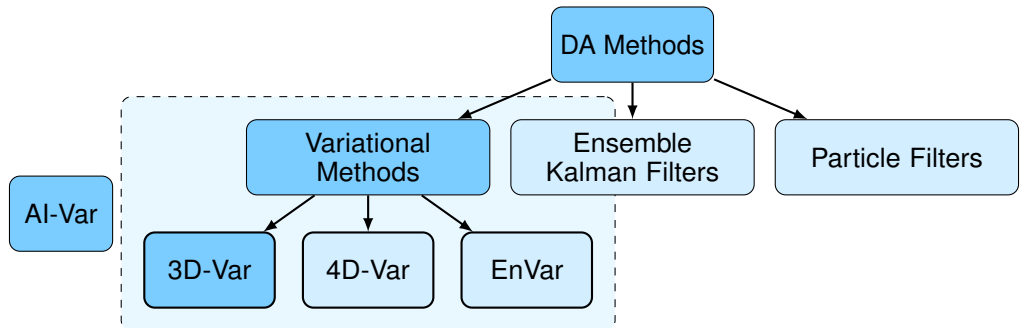

An alternative to hybrid AI approaches, which are currently investigated, is to include observations directly in the AI emulators for NWP, thus, replacing the data assimilation step altogether \citep{Geer2021}. However, current approaches to apply machine learning techniques in data assimilation aim to either substitute parts of the assimilation process with AI-based methods to improve performance \citep{Farchi2021} or to extend the capabilities of current implementations to enhance the quality of the analysis \citep{Buizza2022,Chipilski2023,Qu2024}.

In this paper, however, we strive to replace the classical data assimilation scheme by training a neural network to perform the data assimilation task itself. Specifically, we introduce a concept of AI-based data assimilation which avoids the need of a long time-series of analyses as training set. This approach promises to completely replace a classical data assimilation systems and, thus, opening the field for fully data driven NWP systems in the future. 

Integrating data assimilation into neural networks following variational approaches can build on recent advances in machine learning that have demonstrated the potential of neural networks to learn the complex functional mappings required for optimization, bypassing the computational bottlenecks of traditional methods. By leveraging these capabilities, neural networks can be trained to learn the cost function minimization process inherent in variational data assimilation, offering significant improvements in computational speed and solution accuracy.

In the broader literature, different approaches for solving this problem have emerged in recent years \citep{fan2024artificial}. The relevant approach here is to train neural networks to map input parameters directly to optimal solutions \citep{Liu2022}. Solutions are for example presented for integrating optimization for quadratic problems into neural network architectures \citep[e.g., ][]{Effati2011,Amos2021}. The field of AI-based optimization includes the use of neural networks for guiding and speeding up the minimization such as {\em reinforcement learning} where the neural network learns a policy for optimization \citep{deng2022rlprompt, sukhija2023gradientbased} and the task of {\em meta-training} which employs trained models to help training or minimizing other models \citep{wichrowska2017learned,li2016learning}.  
One key approach for practical applications is function-based learning \citep[e.g., ][]{Briden2024} that trains neural networks using decision-focused loss functions. Overall, integrating neural networks into variational data assimilation leverages recent advances in machine learning to offer a path toward more efficient and accurate state estimation.

Starting with the basic equations of 3D-Var, we present a proof-of-concept for our new AI-based algorithmic data assimilation approach that enables the utilization of deep learning architectures for the assimilation process, where the full AI training can be carried out based on observations and first guess fields only. Being based on the concept of variational data assimilation, we implement the algorithm directly into the training concept of an AI based assimilation system, thus, effectively embodying an \emph{AI-Var}.

The remainder of the paper is structured as follows. Section \ref{sec:methodology} is dedicated to the description of our algorithmic approach, with Section \ref{sec:setup} outlining our experiment framework. Our results are presented in Section \ref{sec:results}, followed by the discussion and conclusions in Section \ref{sec:conclusions}.

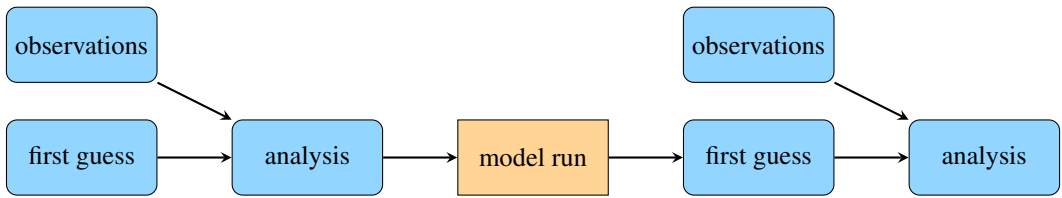
\begin{figure}[ht]
\begin{center}
\begin{tikzpicture}[node distance=2cm]
% Nodes
\node (input2) [fg] {first guess};
\node (input1) [obs, above of=input2, yshift = -0.5cm] {observations};
\node (ana1) [ana, right of=input1, xshift=1cm, yshift=-1.5cm] {analysis};
\node[fill=black!50] (model) [model, right of=ana1, xshift=1cm] {model run};
\node (output) [fg, right of=model, xshift=1cm, yshift=0cm] {first guess};
\node (input3) [obs, above of = output, yshift = -0.5cm] {observations};
\node (ana2) [ana, right of=output, xshift=1cm, yshift=0cm] {analysis};
% Arrows
\draw [arrow] (input1) -- (ana1);
\draw [arrow] (input2) -- (ana1);
\draw [arrow] (ana1) -- (model);
\draw [arrow] (model) -- (output);
\draw [arrow] (input3) -- (ana2);
\draw [arrow] (output) -- (ana2);
\end{tikzpicture}
\end{center}
\caption{\label{fig:da_cycle}
The classical DA cycle, where we replace the step which integrates observations and first guess into an analysis by a direct AI-based assimilation step.}
\end{figure}
%======================================================================================
%
%======================================================================================
\section{Methodology}
\label{sec:methodology}

Current approaches to apply machine learning techniques in data assimilation aim to either substitute parts of the assimilation process with AI-based methods to improve performance and / or to extend the capabilities of current implementations to enhance the quality of the analysis. % Chipilski2023 %\citep[e.g., ][]{Chipilski2023}.
In this paper, however, we develop a concept to replace the classical data assimilation scheme by training a neural network to perform the task. 

%======================================================================================
%
%======================================================================================
\subsection{AI-based Data Assimilation Concepts}
\label{sec:concept}

While previous approaches to utilize AI in data assimilation focused on either extending or substituting aspects of the process, we are here striving to fully integrate data assimilation into a neural network. While machine learning is in general based on the idea of learning relationships from large data sets, we follow a different approach.

\begin{figure}[ht]
\begin{tikzpicture}[node distance=2cm]
\tikzstyle{circlednumber} = [circle, draw, fill=white, inner sep=1pt, font=\sffamily]

% Nodes
\node (input2) [fg] {first guess};
\node (input1) [obs,above of = input2, yshift = -0.5cm] {observations};
\node (ana1) [ana, right of=input2, xshift=1cm] {analysis};
\node (diff) [diff, right of=ana1, xshift=1.5cm, text width=3cm]
% Approach 1
{Approach 1\\minimize difference \\ to train NN};
\node (ana3) [ana, right of=diff, xshift=2cm, yshift=0cm] {AI analysis};
\node (fg2) [fg, right of=ana3, xshift=1cm, yshift=0cm] {first guess};
\node (input3) [obs, above of=fg2, yshift = -0.5cm] {observations};
\node (diff2) [diff, above of=ana3, text width=3cm,yshift=0.2cm]
% Approach 2
{Approach 2\\minimize functional \\ to train NN};

% Grouping box
\begin{pgfonlayer}{background}
    \node [draw,fill=myblue!10, dashed, fit=(input1) (input2) (ana1), inner sep=0.5cm, label=above:Classical Analysis] (group) {};
    \node [draw,fill=myblue!10, dashed, fit=(input3) (fg2) (ana3) (diff2), inner sep=0.5cm, 
    label=above:AI Analysis] (group) {};
\end{pgfonlayer}

% Arrows
\draw [arrow] (input1) -- (ana1);
\draw [arrow] (input2) -- (ana1);
\draw [arrow] (ana1) -- (diff);
\draw [arrow] (ana3) -- node[circlednumber] {1} (diff);
\draw [arrow] (input3) -- (ana3);
\draw [arrow] (fg2) -- (ana3);
\draw [arrow] (ana3) -- node[circlednumber] {2} (diff2);
\end{tikzpicture}
\caption{\label{fig:AI_approaches}
The conventional approach to training a neural network is to minimize the difference between a prescribed analysis and the AI analysis, here Approach 1. The alternative is to minimize a functional as in classical variational minimization, such that the classical analysis is not needed to carry out the minimization. 
}
\end{figure}
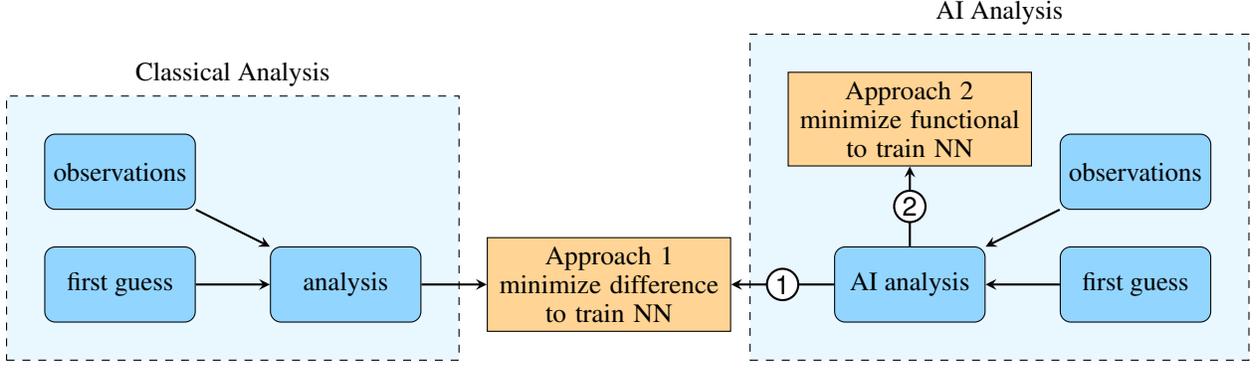

A standard approach of many machine learning applications is to use the AI-based emulators to construct fast and flexible mappings, given input data and the respective desired outcome. For the case of data assimilation, this approach is shown in Figure \ref{fig:AI_approaches} as {\em Approach 1}. In this first approach, a training data set of triples is needed which for each time step consists of {\em observations} $y_{\xi}$, {\em first guess} $x^b_{\xi}$, and {\em analysis} $x^a_{\xi}$, such that we would use the set
\begin{equation}
\label{S1}
{\cal S}_{1} := \big\{ s_{\xi} = (y_{\xi}, x^b_{\xi}, x^a_{\xi}), \quad
\xi =1, ..., n_t \big\}
\end{equation}
with the number $n_t$ of training samples. While this approach appears to be straightforward, it will always require a classical analysis system as basis for the training. Approach 1 is therefore characterized through a strong dependence on the conventional approaches. 

%--------------------------------------------------------------------------------------
%
%--------------------------------------------------------------------------------------

%--------------------------------------------------------------------------------------
%
%--------------------------------------------------------------------------------------
Instead of learning the relationship between input data (i.e., first guess $x^b$, observation $y$) and the desired outcome, i.e. the analysis $x^a$, through training the network on the analysis field as a target, we set the AI model to learn the functional $f(x^b, y) \mapsto x^a$ itself. Therefore, we will make use of the internal loss function of the neural network model. While the loss function commonly only measures the difference between the model output and the provided target in the training, we replace it with a term which penalizes the departure of the outcome (i.e., the analysis) to the first guess plus a term which measures the difference of the model output to the provided observations with weights given by the background and observation error covariance, respectively. Therefore, we end up with a loss function which is exactly the cost function that is commonly used in data assimilation with 3D-Var (compare Figure \ref{fig:da_methods}):
\begin{equation}
\label{eqn:3dvar}
l = (\hbx-\bx^{b})^T \mathbf{B}^{-1} (\hbx-\bx^{b}) + \left({\hat{\mathbf{y}}-\mathbf{y}}\right)^T \mathbf{R}^{-1} \left(\hat{\mathbf{y}}-\mathbf{y}\right)
\end{equation}
where $\delta\hbx = \hbx-\bx^{b}$ is the analysis increment as the output of the neural network with $\hbx$ as the analysis state, $\mathbf{B}$ the background error covariance matrix, $\mathbf{y}$ the observations, $\hat{\mathbf{y}} = H(\hbx)$ the model equivalents of $\hbx$, i.e., the output transformed into observations, and $\mathbf{R}$ the observation error covariance matrix. $\delta\hbx$ is obtained as the model output by minimizing the functional (\ref{eqn:3dvar}). 

We make use of the fact that both variational data assimilation algorithms as well as the AI training algorithm minimize a functional. If we employ the same functional (\ref{eqn:3dvar}) as loss function for training our neural network, we construct the AI-based emulator for data assimilation such that it solves a particular minimization problem -- the data assimilation problem of 3D-Var.

When we train a neural network with the loss function (\ref{eqn:3dvar}), we do not need to prescribe an analysis $x^{a}_{\xi}$ for the input samples $x^{b}_{\xi}$ and $y_{\xi}$. In this case, our training set reduces to 
\begin{equation}
\label{S2}
{\cal S}_{2} := \big\{ s_{\xi} = (y_{\xi}, x^b_{\xi}), \quad
\xi =1, ..., n_t \big\}
\end{equation}
with the number $n_t$ of training samples. This means that we do not need to prescribe an analysis for training. But we need to discuss the role of the first guess fields $x^{b}$ in the whole data assimilation cycle shown in Figure \ref{fig:da_cycle}. 
%--------------------------------------------------------------------------------------
%
%--------------------------------------------------------------------------------------
\begin{wrapfigure}{r}{0.5\textwidth}
\vspace{-5mm}
  \centering
    \includegraphics[width=0.48\textwidth]{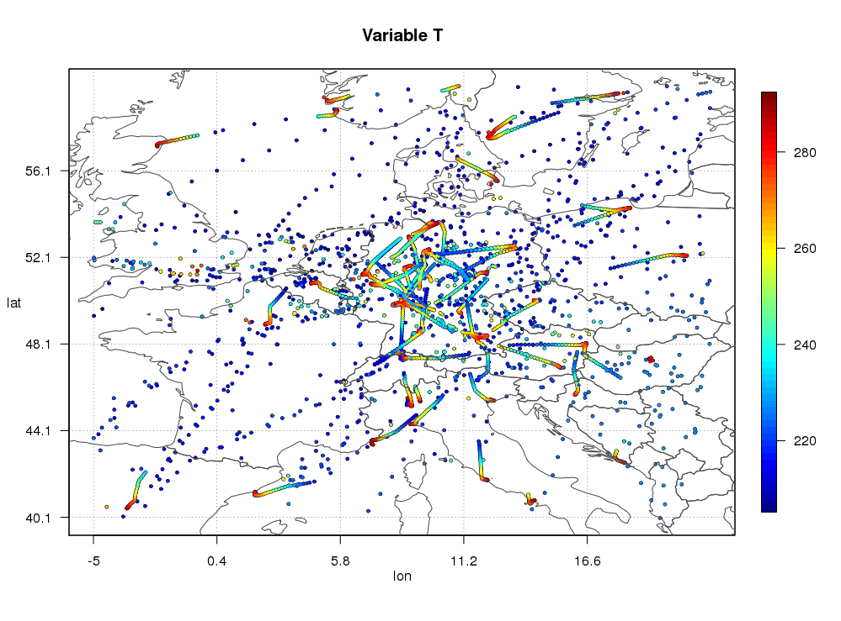}
    \vspace{-5mm}
	\caption{Examples of highly dynamical observations from airplanes over central Europe within one hour. Observations are AIREP upper air temperature in Kelvin.}
 	\label{fig:upper air}
\vspace{-5mm}
\end{wrapfigure}
%--------------------------------------------------------------------------------------
We first note that in classical data assimilation the analysis depends on the observation and the first guess, and of course the first guess is generated from an earlier analysis -- there is a cyclic dependence of the analysis and the first guess through the model propagation in the data assimilation cycle. Usually, when starting a data assimilation cycle, one would try to generate an analysis with a preliminary first guess followed by a spin-up period for the cycle. 

The challenge is even stronger for the AI-based analysis algorithms, since we do not solve the minimization problem in each step, but we train an emulator to solve this minimization problem -- and the training is carried out by solving many minimization problems. However, when a reasonable first guess is available, we can use it for training. When no first guess is available, we might work with $x^b=0$ initially. When a first version of an analysis algorithm after training with $x^b=0$, one can construct a better analysis algorithms based on the available first guess fields calculated from available analyses in an iterative manner.  Here, we will focus on the analysis step itself, showing that indeed the above approach is feasible, further refinements through iterations will be a topic of future research. 
 
Approach 1 is using the {\em supervised learning} technique in the sense that the target solution $x^{a}$ is prescribed and the AI-based analysis is trained to get as close to the target as possible. Unsupervised learning algorithms carry out their tasks without such prescribed targets or labels, respectively. Our Approach 2 is a self-supervised learning approach that is a sub-category of unsupervised learning algorithms. For further details on these categorizations we refer to \citet{James2013,Bishop2006, Goodfellow2016,Hastie2001}.
For our conceptional tests we have chosen three basic configurations: 
\begin{enumerate}
\item[a)] a {\em one-dimensional} idealized configuration,
\item[b)] a {\em two-dimensional} idealized configuration,
\item[c)] a {\em real-world} test case (2m temperature analysis).
\end{enumerate}

In addition, we will investigate two basic {\em observational setups} that are encountered in real world data assimilation problems: 
\begin{enumerate}
\item[(1)] the observations are available in a {\em fixed} configuration - static observation locations,
\item[(2)] the places where observations are taken are {\em flexible} - dynamic observation locations. 
\end{enumerate}
Both observational setups reflect real-world problems of high relevance. For example, our real-world case when assimilating SYNOP two-meter temperature measured at standard operational weather stations falls into category (1). When we want to assimilation measurements of commercial airplanes as shown in Figure \ref{fig:upper air}, which are important observations of toady's global or regional forecasting systems, the observational setup would be of category (2). 

%======================================================================================
%
%======================================================================================
\subsection{Assimilation of fixed-position observations}
\label{sec:fixed}

For testing the feasibility of the new Approach 2 we have used PyTorch, the well-known open-source machine learning library developed by Facebook's AI Research lab. We employed PyTorch version 2.2.0 for neural network training and experimentation, see \cite{pytorch2024}.
\begin{comment}
PyTorch. (2024). PyTorch: An open-source machine learning library. Retrieved from https://pytorch.org/
\end{comment}
Pseudo code for the core procedure is given in the Appendix \ref{sec:appendix_psuedo_code}. 
Conceptually, we need to provide (a) {\em input fields}, (b) implement our choice of a neural architecture with input and output dimensions and (c) define a loss function which the minimizer of pytorch can use. Further, we need to provide a training and evaluation data set for the minimization framework to run. 

In the basic configuration a) defined in Section \ref{sec:concept}, where observations are at fixed locations, we can calculate observation equivalents from model fields based on a fixed set of parameters given by the observation locations. The observation operators usually carry out an interpolation of model fields to observation locations, often in combination with further physical processes such as the transmission of radiation. Here, we have chosen to stay with a simple framework where the model field consists of one variable and we measure the model field in a selection of model grid locations. In this case, the observational parameters are a list of indices of the model grid points where the values are measured. The observations are modeled by a simple array of scalar values. In the case of $m$ temperature observations at particular locations, this array will consist of $m$ values, for which the model equivalents can be obtained taking the $m$ indices of the model grids where the measurement took place, and evaluating the corresponding field values. 

% %--------------------------------------------------------------------------------------
% % FIGURE - 1D EXAMPLES
% %--------------------------------------------------------------------------------------
% \begin{wrapfigure}{rt}{0.5\textwidth}
% \vspace*{-0mm}
%   \centering
%     \includegraphics[width=0.48\textwidth]{1D_functions.pdf}
%     \vspace{-5mm}
% 	\caption{Examples of the functions used in training the 1D version of the data assimilation model.}
%  	\label{fig:1D_functions}
%     \vspace{-10mm}
% \end{wrapfigure}
% %--------------------------------------------------------------------------------------

(a) For PyTorch, the input is a combined array of values $(x^{b},y)$ where $x^{b}$ is the first guess field transformed into the inherent PyTorch tensor structure, and $y$ is the array of observations. With $n$ gridpoints and a field of one variable only, the tensor $(x^{b},y)$ has $n+m$ values. 

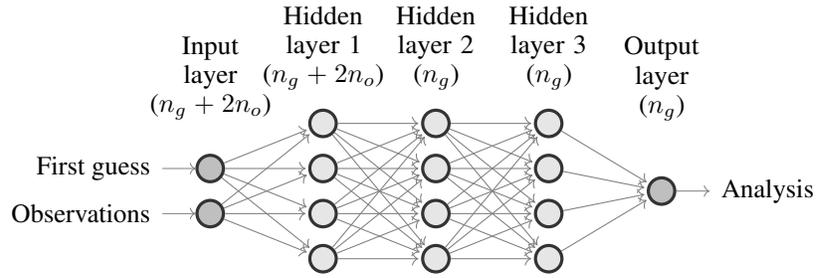
\begin{wrapfigure}{r}{0.65\textwidth}
\vspace{-12mm}
\centering
\begin{tikzpicture}[scale=0.6,shorten >=1pt, ->, draw=black!50, node distance=\layersep]

\tikzstyle{input neuron}=[very thick, circle, draw=black!80, fill=gray!50, minimum size=10pt, inner sep=0pt]
\tikzstyle{hidden neuron}=[very thick, circle, draw=black!80, fill=gray!20, minimum size=10pt, inner sep=0pt]
\tikzstyle{output neuron}=[very thick, circle, draw=black!80, fill=gray!50, minimum size=10pt, inner sep=0pt]
\tikzstyle{annot} = [text width=6em, text centered]

% Define the distance between layers
\def\layersep{2.5cm}

% Draw the input layer nodes
%\node[input neuron, pin={[pin edge={->}]left:First guess}] (I-1) at (0,-1.5) {};
\node[input neuron, pin={[pin edge={<-}]left:First guess}] (I-1) at (0,-1.5) {};
\node[input neuron, pin={[pin edge={<-}]left:Observations}] (I-2) at (0,-2.5) {};
% Draw the hidden layer 1 nodes
\foreach \name / \y in {1,...,4}
    \path[yshift=0.5cm]
        node[hidden neuron] (H1-\name) at (\layersep,-\y cm) {};

% Draw the hidden layer 2 nodes
\foreach \name / \y in {1,...,4}
    \path[yshift=0.5cm]
        node[hidden neuron] (H2-\name) at (2*\layersep,-\y cm) {};

% Draw the hidden layer 3 nodes
\foreach \name / \y in {1,...,4}
    \path[yshift=0.5cm]
        node[hidden neuron] (H3-\name) at (3*\layersep,-\y cm) {};

% Draw the output layer nodes
\foreach \name / \y in {1,...,1}
    \path[yshift=0.5cm]
        node[output neuron, pin=right:Analysis] (O-1) at (4*\layersep,-2.5cm) {};

% Connect every node in the input layer with every node in the hidden layer 1
\foreach \dest in {1,...,4}
        \path (I-1) edge (H1-\dest);
\foreach \dest in {1,...,4}
        \path (I-2) edge (H1-\dest);

% Connect every node in the hidden layer 1 with every node in the hidden layer 2
\foreach \source in {1,...,4}
    \foreach \dest in {1,...,4}
        \path (H1-\source) edge (H2-\dest);

% Connect every node in the hidden layer 2 with every node in the hidden layer 3
\foreach \source in {1,...,4}
    \foreach \dest in {1,...,4}
        \path (H2-\source) edge (H3-\dest);

% Connect every node in the hidden layer 3 with every node in the output layer
\foreach \source in {1,...,4}
    \foreach \dest in {1,...,1}
        \path (H3-\source) edge (O-\dest);

% Annotate the layers
\node[annot,above of=H1-1, node distance=1cm] (hl1) {Hidden layer 1\\($n_g+2n_o$)};
\node[annot,above of=H2-1, node distance=1cm] (hl2) {Hidden layer 2\\($n_g$)};
\node[annot,above of=H3-1, node distance=1cm] (hl3) {Hidden layer 3\\($n_g$)};
\node[annot,above of=I-1, node distance=1.2cm] {Input\\layer\\($n_g + 2n_o$)};
\node[annot,above of=O-1, node distance=1.5cm] {Output\\layer\\($n_g$)};

\end{tikzpicture}
\caption{Network architecture for the idealized cases. Below the layer names, the dimension of each layer is provided. Here, $n_g$ is size of the grid space (i.e., $n_g=n_x$ for the 1D case and $n_g=n_x\cdot n_y$ for the 2D case) and $n_o$ is the number of observations.}
\label{fig:1D_network}
\vspace{-12mm}
\end{wrapfigure}

(b) The neural architecture has been adapted to the three cases a)-c) and will be described in more detail in the sections below. We basically used dense networks for our test cases as shown in Figure \ref{fig:1D_network}. 

(c) The cost function can be provided in PyTorch as a function based on its own data structures, the PyTorch tensors. The tensorial structure allows us define the complex loss function (\ref{eqn:3dvar}) as an implementation of the Mahalanobis distance (see \cite{Mahalanobis1936}) easily even in the multi-dimensional case. 

Finally, we need to define a training data set. For the test configurations a) and b) we have chosen set of prescribed {\em true} functions for simulation, used a random number generator to select a set of indices to fix observation locations, and calculated artificial observations based on this data set. 
We have split the data set into a training data set and an evaluation data set.   
With these ingredients the setup of the training loop based on a data loader, an optimizer, and the optimization criterion is straightforward in PyTorch. 

%======================================================================================
%
%======================================================================================
\subsection{Assimilation of flexible-position observations}
\label{sec:flexible}

For the case of flexible-position observations a simple approach is to add the location of the input fields to the input vectors given to the AI-based system. This means that we have a parameter vector $p_{\xi}$ which depends on the sample index $\xi$ which together with the first guess $x^{b}_{\xi}$ and the observations $y_{\xi}$, $\xi=1,...,N$ with $N$ being the number of training samples is given as input to the system. The set of training samples now takes the form 
\begin{equation}
\label{S3}
{\cal S}_{3} := \big\{ s_{\xi} = (y_{\xi},  p_{\xi}, x^b_{\xi}), \quad
\xi =1, ..., N \big\}, 
\end{equation}
where in the simplest possible case of linear indices the dimension of $y_{\xi}$ and $p_{\xi}$ is identical. Depending on the particular format of indexing chosen, of course $p_{\xi}$ can take tensorial form. The PyTorch system is able to deal with any of these, as long as the structures are transformed into appropriate PyTorch tensors. 

We remark that adding the locations to the system increases the computational complexity of the learning task considerably. Initially, we naively added the indices, but in these cases our training tasks failed. As described in \citep{Liu2022}, for the case where the order of the observations does not change the minimization problem, deep sets and {\em permutation invariant} neural architectures can be used to reduce the complexity of the training task. Here, we employed a simple approach to such methods by ordering the observations based on their location in the grid, which leads to a uniform and permutation invariant structure for all permutations of observations in the data assimilation task. With this approach we were successful to achieve meaningful results. 

%--------------------------------------------------------------------------------------
% FIGURE - NETWORK ARCHITECTURE
%--------------------------------------------------------------------------------------
%--------------------------------------------------------------------------------------

%======================================================================================
% SECTION - EXPERIMENT SETUP
%======================================================================================

\section{Experiment setup}
\label{sec:setup}

In this section, we describe the data used in the three test cases for our data assimilation learning approach. We employ two idealized cases, one- and two-dimensional, respectively, as well as a real world example.

\subsection{1D setup}
\label{sec:1dsetup}

The first case examines our data assimilation learning algorithm using data from one-dimensional functions. We use two different basic functions to generate these data.

In our setup, the x-axis values represent a linearly spaced sequence of $n_x=50$ values in the interval  \( [0,2\pi[ \). With \( f \sim \text{Uniform}(0.5, 1.5) \) a random forcing value and \( x_{t} = x - t \cdot \Delta t \) the shifted $x$ values along the x-axis for \(n_t\) number of time steps, we define \textit{sinusoidal curves} as \(z_t^{\text{s}}=\sin{x_t}\) and \textit{parabolas} using \(z_t^{\text{p}}=f\cdot{x_t}^2\). %and \textit{cubic curves} with \(z_t^{\text{c}}=f\cdot\left(\Tilde{z}^3-\Tilde{z}^2\right)\).
The data set is constructed such that, at each time step, the data is randomly chosen from each of the two types of curves. Examples of the resulting curves are shown in the results section in Figure \ref{fig:1D_examples}.

% %--------------------------------------------------------------------------------------
% % FIGURE - 2D EXAMPLES
% %--------------------------------------------------------------------------------------
% \begin{wrapfigure}{r}{0.46\textwidth}
% \vspace{-2mm}
%   \centering
%     \includegraphics[width=0.45\textwidth]{2D_fields.pdf}
%     \vspace{-2mm}
% 	\caption{Exemplary fields for the idealized 2D test case showing observation locations for the static (white) and dynamic (black) setup with $n_o=5$.}
%     \vspace{-5mm}
%  	\label{fig:2D_fields}
% \end{wrapfigure}
% %--------------------------------------------------------------------------------------

To test the functionality of the data assimilation approach in the 1D idealized case, we apply the following experiment setup. First, we sample our observations from the training data set of true model states. We choose the first guess state to be zero and, then force the model to reconstruct the true fields by assimilating the observations. In this case, to give the observations an appropriate impact reflecting a large uncertainty of the background, we introduce artificial weights $w_b=0.001$ for the background term and $w_o=0.999$ from the observation term into the loss function.

In order to investigate the performance of our model for the data assimilation purposes, we employ two common scenarios - a static and a dynamic layout of observation locations. We then investigate the robustness of the model for the different scenarios with respect to three hyperparameters. First, we train models with different number of observations $n_o=5,10,15,20$ to represent cases from a sparsely to densely observed space. Second, we use different Gaussian kernel widths sigma ($s=0.5,1.,2.,4.$) for the B-matrix to account for various radii of influence for the observations on the state space. Last, we train the model with different training sample sizes of $n_t=250,500,1000,2500$. For all experiments we use $100$ independent samples for evaluation. The samples are drawn without replacement using the different random forcings $f$.

Our architecture consist of a simple feed forward neural network with three hidden layers as depicted in Figure \ref{fig:1D_network}. The input of the neural network is the first guess field (size $n_x$) as well as the observations (size $n_o$) and the observation indices in the spatial dimension (also size $n_o$), made permutation invariant as described in the methodology section. For the implementation of the algorithm, we use the python package PyTorch. The used optimizer is Adam and the learning rate is set to 0.001.

\subsection{2D setup}
\label{sec:2dsetup}

To systematically investigate the functionality of our approach for a more realistic data assimilation problem, we employ a 2D setup using trigonometric functions which is designed similar to the 1D case. Here, we use a 2D grid described by the coordinates $x$ and $y$. Then, we define the factors $f_x$ and $f_y$ determining the different displacement in $x$ and $y$ directions for the samples through $c_x=f_x \cdot \pi \cdot t/n_t$ and $c_y=f_y \cdot \pi \cdot t/n_t$, respectively. The 2D field $z$ is then calculated by $z=\sin\left[ (\pi / 5) \cdot (x - c_x[t])\right] \cdot \cos\left[(\pi / 8) \cdot (y - c_y[t]) \right]$. Finally, the array $z$ is permutated along the sample axis. The idealized 2D experiments follow the exact same setup as for the 1D case (cf. section \ref{sec:1dsetup}) including the model architecture and parameter settings, including the {\em permutation invariance} described above. Examples for the resulting fields are shown in the results section in Figure \ref{fig:2D_examples}.

%--------------------------------------------------------------------------------------
% FIGURE - 1D VALIDATION LOSS 
%--------------------------------------------------------------------------------------
\begin{figure}[t]
\vspace{0mm}
  \centering
    \includegraphics[width=0.99\textwidth]{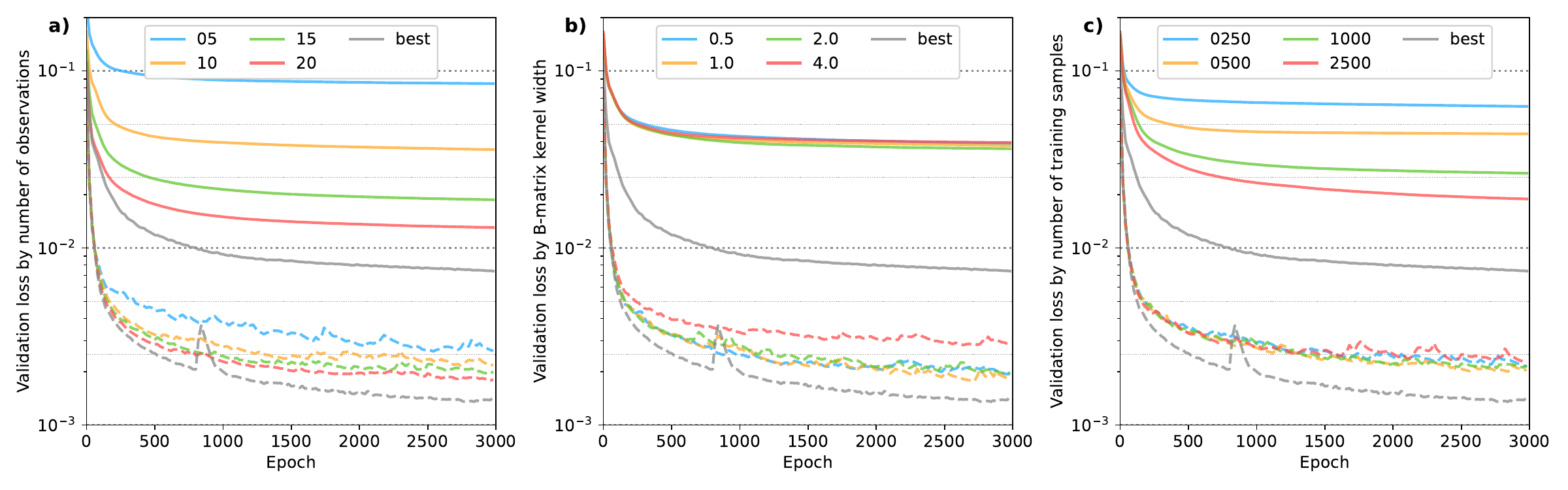}
	\caption{Validation loss curves for the 1D idealized case for the two scenarios - static (dashed lines) and dynamic (solid lines) observation locations. The curves represent the mean validation loss over the group of respective training runs with the same a) number of observations, b) B-matrix kernel width, and c) number of training samples. Therefore, each line is the average over 16 data sets, respectively.}
 	\label{fig:1D_loss}
\vspace{0mm}
\end{figure}
%--------------------------------------------------------------------------------------

\subsection{Real case setup}
\label{sec:realsetup}

In order to evaluate the potential of our AI-Var approach in a real world setting, we add another experiment setup where we apply our approach to assimilate 2-meter temperature (T2M) observations from synoptic observation sites into T2M fields from an NWP model. This setup corresponds to our 2D idealized static observation location experiment. As gridded data, we choose hourly T2M fields from DWD's COSMO-REA6 regional reanalysis \citep{Bollmeyer2015}. Instead of using the full data set, a subdomain of $20 \times 25$ grid points (at 6km horizontal resolution) over central Germany is selected to reduce the complexity. While conventional observations were assimilated in the generation process of the reanalysis, T2M measurements have only been introduced indirectly via a soil moisture reanalysis once a day. Therefore, analyses of the reanalysis T2M estimates showed considerable biases, RMSE and correlation mismatches \citep{Keller2021}.

The observational data comes from DWD's climatological archive where synoptic measurements are stored for German stations. The subdomain includes 25 stations with hourly observations. Times where data from any station are missing have been removed from the whole data set.

Our test period comprises four years (2010-2013) where we only use the summer months to concentrate on Northern hemisphere summer. The data set comprises 8658 samples which are randomly permutated along the time axis. We retain 500 data points as validation data set and use the remaining 8158 samples for training with a batch size of 4096. We employ the same setup as for the 2D case except for a few tuning choices. (1) To better account for the complexity of the underlying dynamics, we provide the model surface height (which is closely connected to T2M) as a 2D field as well as time of day as a scalar to the model. (2) We adapt the neural network architect by increasing the number of neurons in hidden layer 1 and 2 to $n_g\times 4$ and $n_g\times 2$, respectively. 

(3) We still use a Gaussian B-matrix which might not be the best choice but should suffice in this proof-of-concept, but we set the B-matrix kernel width to $0.2$ and the observation error to $0.1$.

In a secondary test setup, we withhold observations to allow for an independent verification data set. In this cross-validation setup, we perform five model training runs with $n_o=20$ by each time keeping a subset of five observations from the original data set for evaluation in such a way that we have an independent model for each observation.

%======================================================================================
% SECTION - RESULTS
%======================================================================================

%--------------------------------------------------------------------------------------
% FIGURE - 1D EXAMPLES
%--------------------------------------------------------------------------------------
\begin{figure}[t]
\vspace{0mm}
  \centering
    \includegraphics[width=0.99\textwidth]{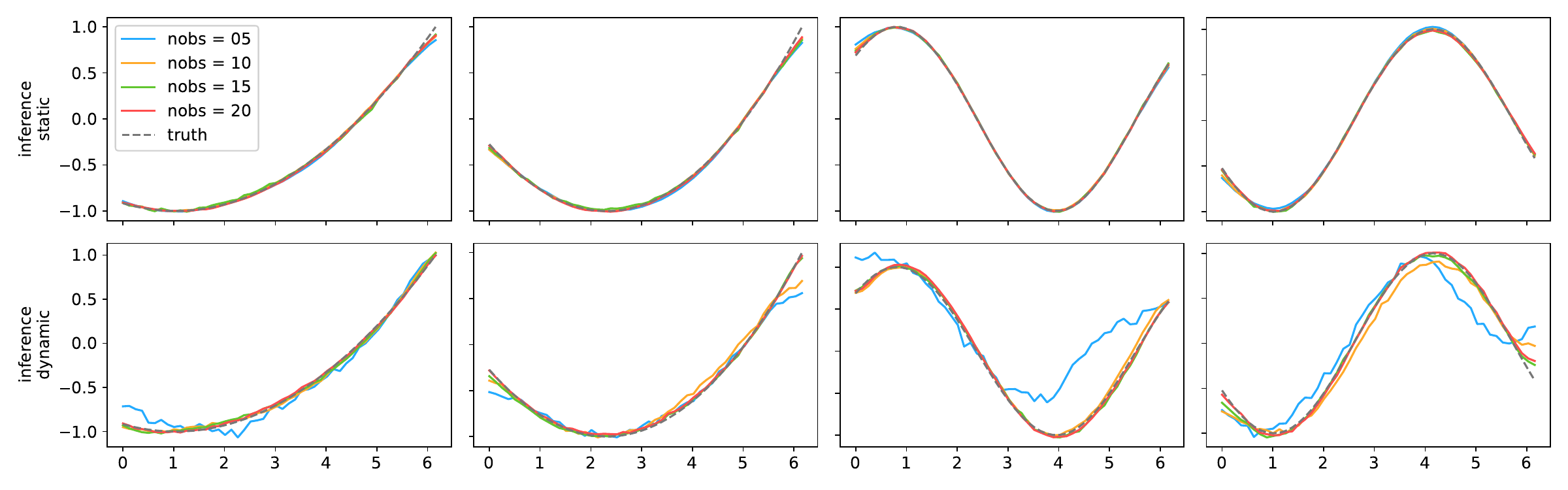}
	\caption{Examples of the 1D idealized experiments. The plots depict the 1D fields for four inference samples of the validation data set (columns) from the static (top row) and the dynamic observation locations (bottom row). The grey dashed lines denote the respective true state.}
 	\label{fig:1D_examples}
\vspace{0mm}
\end{figure}
%--------------------------------------------------------------------------------------

\section{Results}
\label{sec:results}

In this section, we show the results from our proof-of-concept implementation of an AI-based data assimilation algorithm. We first present the results for the idealized 1D and 2D cases and conclude this section with a simple real-world 2D example.

\subsection{Idealized 1D case}
\label{sec:results_1D}

We first look at the loss curves of the training of our data assimilation AI model. Specifically, we are interested in the validation loss, as the aim of the approach is to be applied to independent data. The three plots in Figure \ref{fig:1D_loss} show the average validation loss for all training runs with a specific parameter setting as described in Section \ref{sec:1dsetup}, i.e. each colored line represents the average over 16 different models. The dashed lines are determined using a static observation setup whereas the solid lines indicate the average over models with dynamic observation positions. The grey lines indicate the best performing model for both cases.

The training loss (not shown) as well as the validation loss converges for all models. The comparison (Fig. \ref{fig:1D_loss}) further shows that the models with static observation locations perform better than the ones with dynamic observation locations (i.e., the former are exhibiting a lower average validation loss). This was expected as the static structure allows for the neural network to better learn the characteristic of the underlying fields with respect to the provided observations.

Looking at the parameter "number of observations" (Fig. \ref{fig:1D_loss}a), we find a strong deviation in performance between the models especially for the dynamic setup with 20 observations showing a validation loss nearly one order of magnitude lower than the models with only 5 observations. Remembering that our 1D space has 50 grid points, the case with 20 observations prescribes already nearly half of the state space, thus, significantly reducing the number of grid points to be determined compared to the 5 observation case. For the static observation locations, the validation loss varies only slightly with respect to the number of observations (between 0.002 and 0.004 at the end of the training), again, with 20 observations exhibiting the lowest validation loss.

For the various B-matrix kernel width settings (Fig. \ref{fig:1D_loss}b), we find that the parameter does only exhibit a very small impact on the performance of the model with dynamic observation locations. For the static setup however, we find that while the lower settings have similar validation losses, a value of $4.0$ seems to imply a too strong smoothing which increases the validation loss.

The number of training samples seems to be a crucial parameter for the performance of the models with dynamic observation locations. Figure \ref{fig:1D_loss}c shows that there is strong variation among the different settings with the largest number of samples (2500) having the smallest validation loss. This is founded in the ability of the model to generalize from the changing observation network layout by more samples and, thus, a larger number of different combinations of observation locations. The impact on the static observation location setup is, however, only small, as a limited number of samples is sufficient for the model to understand the underlying structure of the fields.
%--------------------------------------------------------------------------------------
% FIGURE - 2D VALIDATION LOSS
%--------------------------------------------------------------------------------------
\begin{figure}[t]
\vspace{0mm}
  \centering
    \includegraphics[width=0.99\textwidth]{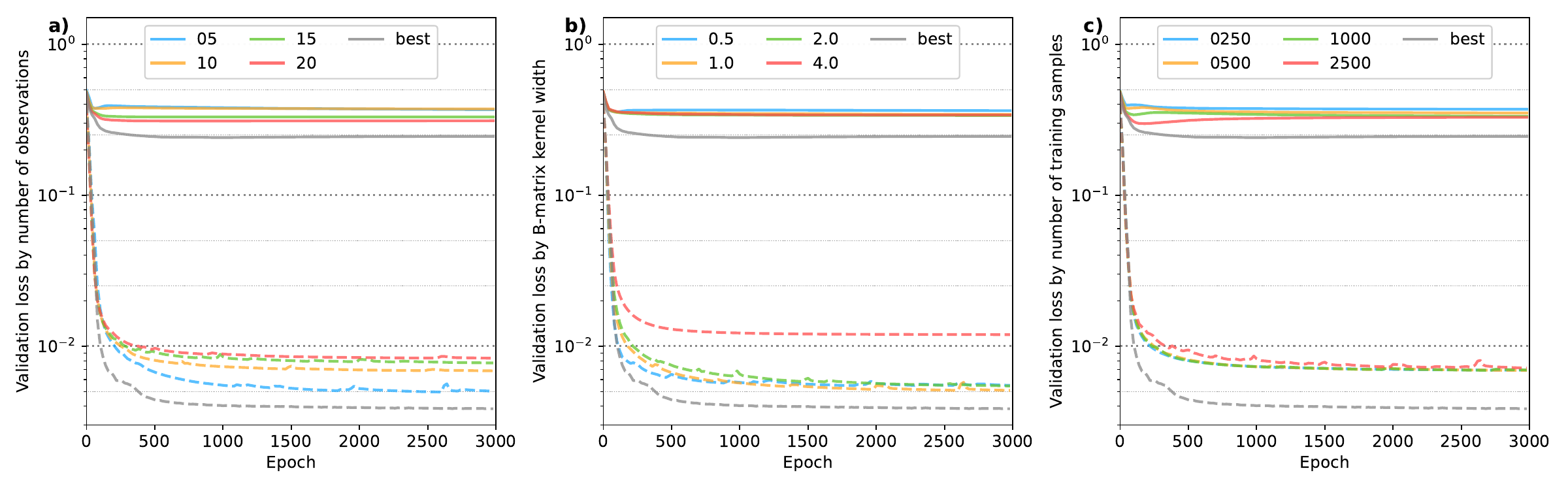}
	\caption{Validation loss curves for the 2D idealized case for the two scenarios - static (dashed lines) and dynamic (solid lines) observation locations. The curves represent the mean validation loss over the group of respective training runs with the same a) number of observations, b) B-matrix kernel width, and c) number of training samples. Therefore, each line is the average over 16 data sets, respectively.}
    \vspace{-5mm}
 	\label{fig:2D_loss}
\end{figure}
%

%--------------------------------------------------------------------------------------
% FIGURE - 2D RMSE
%--------------------------------------------------------------------------------------
\begin{wrapfigure}{r}{0.6\textwidth}
\vspace{-5mm}
  \centering
    \includegraphics[width=0.59\textwidth]{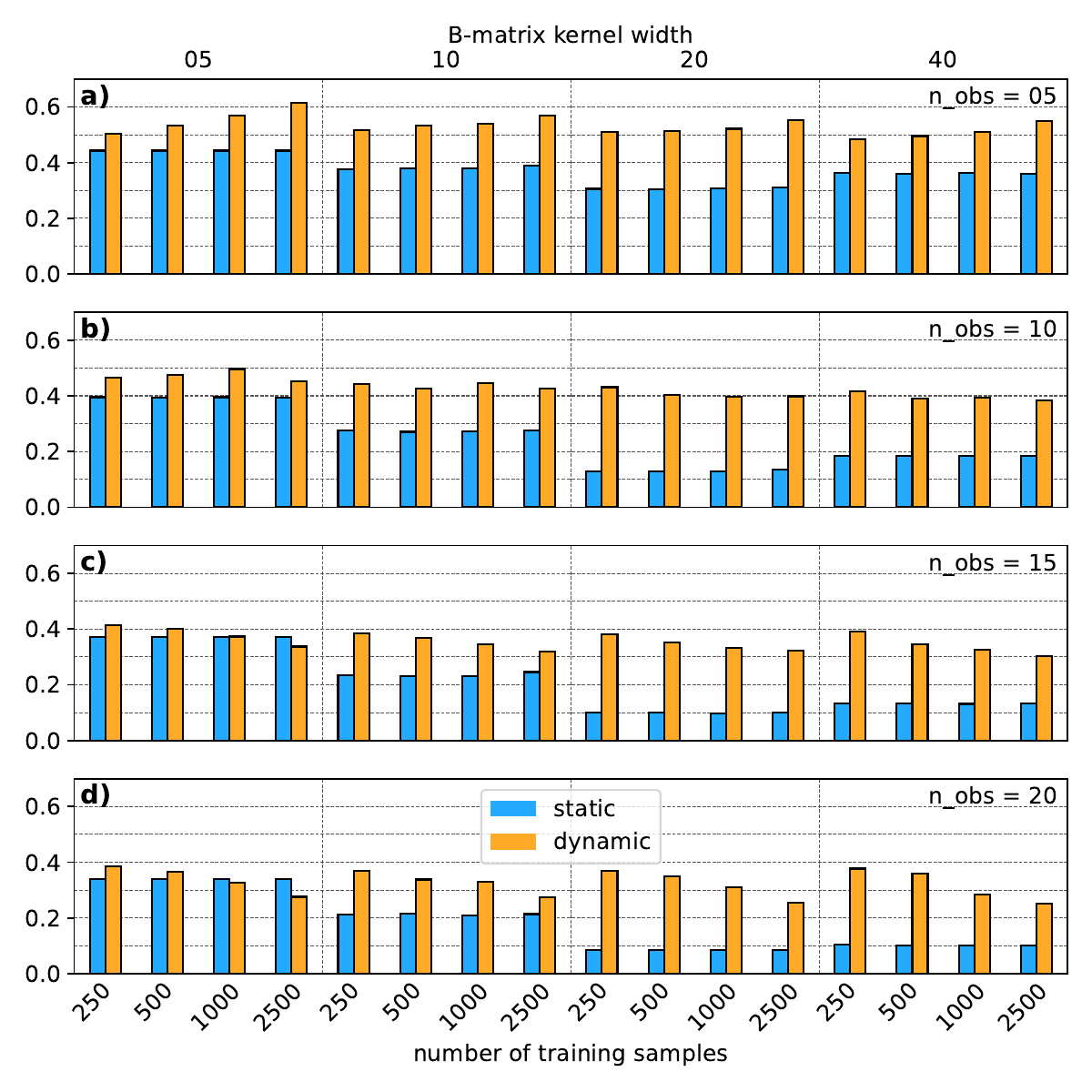}
    \vspace{-2mm}
	\caption{RMSE for inference against truth over all 100 validation samples and all grid points for the static (blue) and dynamic (yellow) observation locations. The four plots represent the results for the models based on a) 5, b) 10, c) 15 and d) 20 observations.}
 	\label{fig:2D_rmse}
\vspace{-8mm}
\end{wrapfigure}
%--------------------------------------------------------------------------------------

To take a closer look at the generated fields, Figure \ref{fig:1D_examples} shows four exemplary samples from the validation data set (columns) - two parabolas (left) and two sinus curves (right) with the respective inference results for all four numbers of observations $n_o$ with $n_s=2500$ and $s=2.0$. In the top row, the curves are reconstructed with the static observation locations setup whereas the bottom row depicts the results from the dynamic observation locations.

The results indicate that our AI-Var approach is able to very closely reconstruct the true state for the static case for the given parameter settings independent of the number of observations. For the dynamic case, we find that for $n_o={15,20}$, the model is also nearly identical to the truth except for some data points close to the boundaries when there are no nearby observations. While the estimates slightly deviate from the truth for $n_o=10$, we find that for $n_o=5$, the inference is not very smooth anymore and shows larger discrepancies from the original state. This indicates that for a less dense observation network, a larger training data set can compensate this issue for the static case, but it seems to be much more complicated to achieve good results for the dynamic case.

\subsection{Idealized 2D case}
\label{sec:results_2D}

We now anaylze the results with respect to the idealized 2D case. Figure \ref{fig:2D_loss} depicts the average validation loss over the training epochs for all training runs similar to the 1D case (cf. Fig. \ref{fig:1D_loss}). As expected, we find that independent of the hyperparameter settings, the static observation locations exhibit lower validation losses compared to the dynamic setup. This difference is considerably increased in comparison to the 1D case which can be attributed to the more complex 2D situation.

%--------------------------------------------------------------------------------------
% FIGURE - 2D EXAMPLES
%--------------------------------------------------------------------------------------
\begin{figure}[t]
\vspace{0mm}
  \centering
    \includegraphics[width=0.99\textwidth]{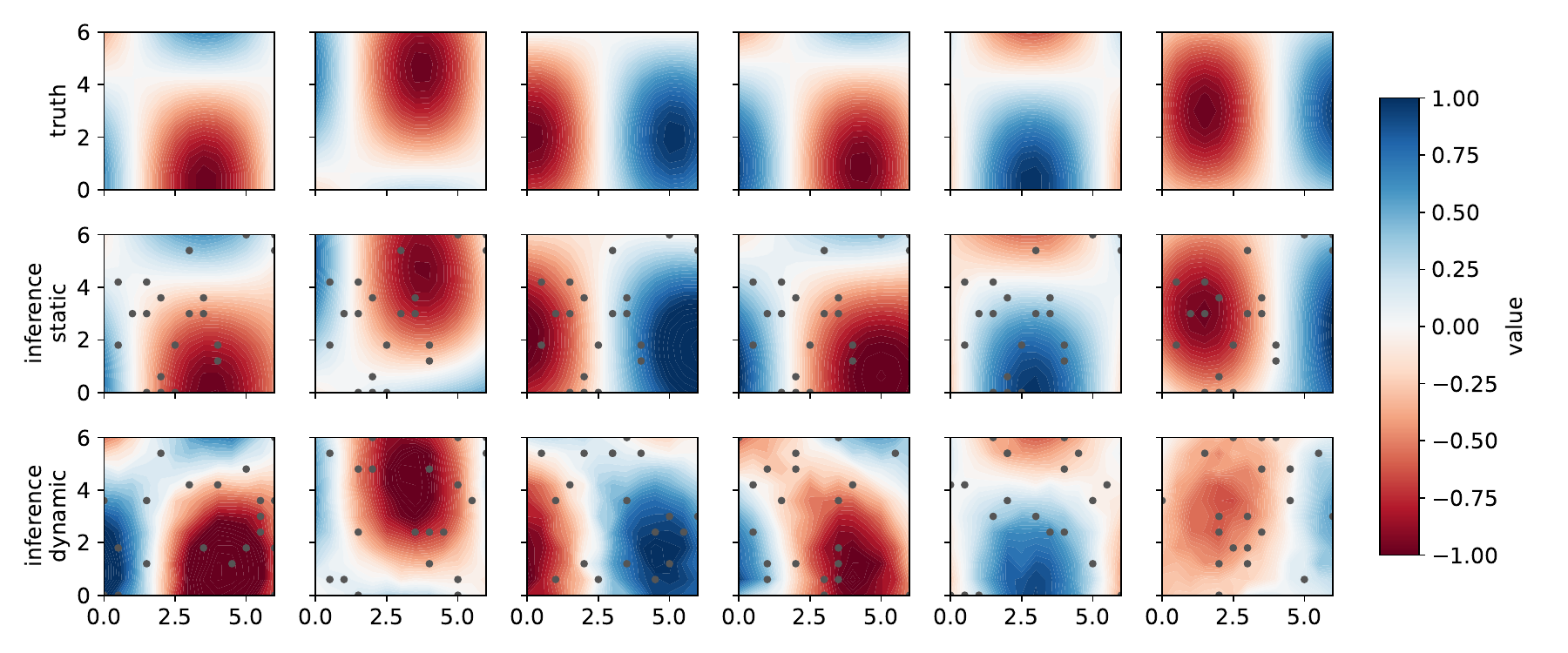}
	\caption{Examples of the 2D idealized experiments. The plots depict the 2D fields for six samples of the validation data set (columns) from the truth (top row), the inference (i.e.\ the AI based analysis) from the static (middle row) and the dynamic observation locations (bottom row). The grey dots denote the respective observation locations.}
 	\label{fig:2D_examples}
\vspace{0mm}
\end{figure}
%--------------------------------------------------------------------------------------

In Figure \ref{fig:2D_loss}a, we find that the impact of the number of observations is far more pronounced for the static case than for the dynamic observation locations. For the latter, we see that the best performing models are those with the largest number of observations. However, for the static locations, this effect is reversed with lower number of observations showing a lower validation loss. For the B-matrix kernel width (Figure \ref{fig:2D_loss}b), most of the respective models exhibit a similar performance except for a width of 4 for the static case where the strong smoothing appears to hinder the ability of the neural network to estimate the observations, thus leading to a considerably higher validation loss.
%--------------------------------------------------------------------------------------
% FIGURE - T2M LOSS
%--------------------------------------------------------------------------------------
\begin{wrapfigure}{r}{0.41\textwidth}
\vspace{-3mm}
  \centering
    \includegraphics[width=0.40\textwidth]{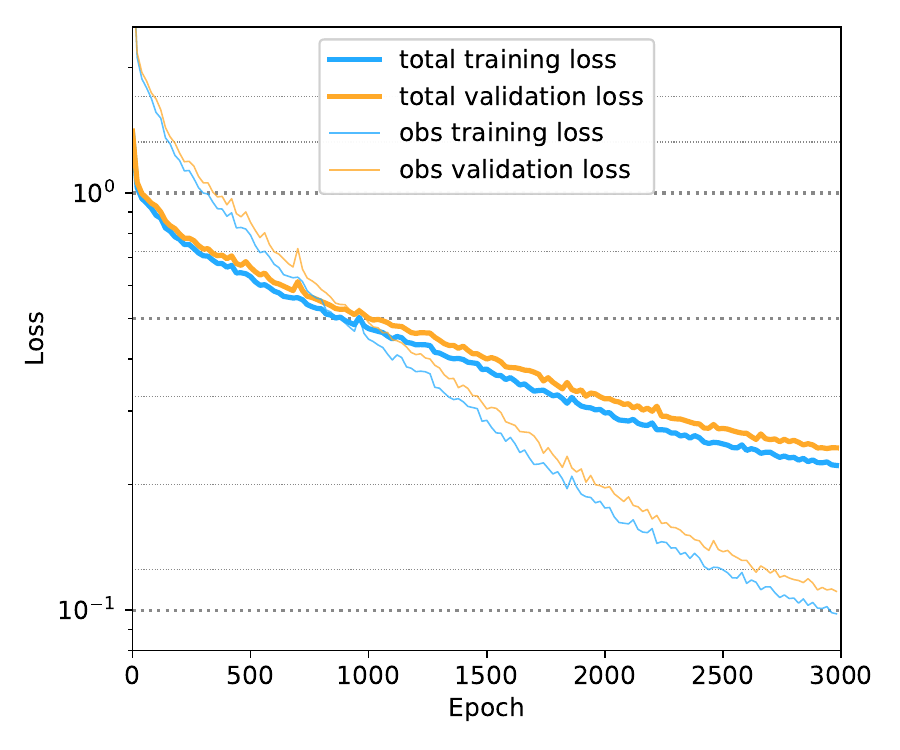}
	\caption{Loss function for the real test case. The thick lines depict the total loss whereas the thin lines are the loss coming from the observation term. Blue indicates training loss and yellow denotes validation loss.}
 	\label{fig:t2m_loss}
\vspace{-4mm}
\end{wrapfigure}
%--------------------------------------------------------------------------------------
For the training sample size (Figure \ref{fig:2D_loss}c), we find that the results are similar for the static observation locations, indicating that such a setup is robust also against smaller sample sizes. For the dynamic locations, we find that larger sample sizes lead to smaller validation losses. However, especially for 2500 samples, we see that the average validation loss decreases until a minimum is reached (around epoch 150) and then increases until convergence is reached. As we look at the validation loss here, this indicates that while larger sample sizes can improve the model quality, there is a considerable risk of overfitting the model towards the training samples.

To further investigate the performance of the AI-based DA approach in a 2D idealized setting, we perform the inference on the validation data set with all models and compare the RMSE of the estimated fields to the truth. Figure \ref{fig:2D_rmse} shows the average RMSE over all grid points and all samples for the static (blue) and dynamic (yellow) observation locations setup. The four plots indicate the different number of observations as these are usually given in a real world setting. In general, we find that the model for the static setup is much less insusceptible towards the B-matrix kernel width whereas the models for the dynamic locations perform much better with larger B-matrix kernel widths. For the latter, a setting of 2 seems to be the best of the choices for this case. In comparison, the number of training samples has a much larger influence on the performance of the models for static locations than for the dynamic locations. The influence on the former also depends on the number of observations. With 5 observations, smaller samples produce higher RMSEs whereas for 20 observations, the best results can be achieved with larger number of samples. This clearly indicates that the model's tendency for overfitting is dependent on the complexity of the situation (i.e., the number of observations).

Six examples of the reconstruction from the validation data set for both the dynamic and the static case can be found in Figure \ref{fig:2D_examples}. The inference is performed for the models with $2500$ training samples, $20$ observations and a B-matrix kernel width of $2.0$. The top row depicts the original fields, wheres the center and bottom row show the reconstruction for the static and dynamic observation locations, respectively. The dots denote the observation location for each example. We find that the models are able to quite reasonably reproduce the 2D structure of the truth. While in the static case, the differences between truth and reconstruction are only marginal, some deviations become apparent for the dynamic case.

In summary, the results for 2D idealized experiments are very promising indicating the potential for the application in a real test case.

%--------------------------------------------------------------------------------------
% FIGURE - T2M EXAMPLES
%--------------------------------------------------------------------------------------
\begin{figure}[t]
\vspace{0mm}
  \centering
    \includegraphics[width=0.99\textwidth]{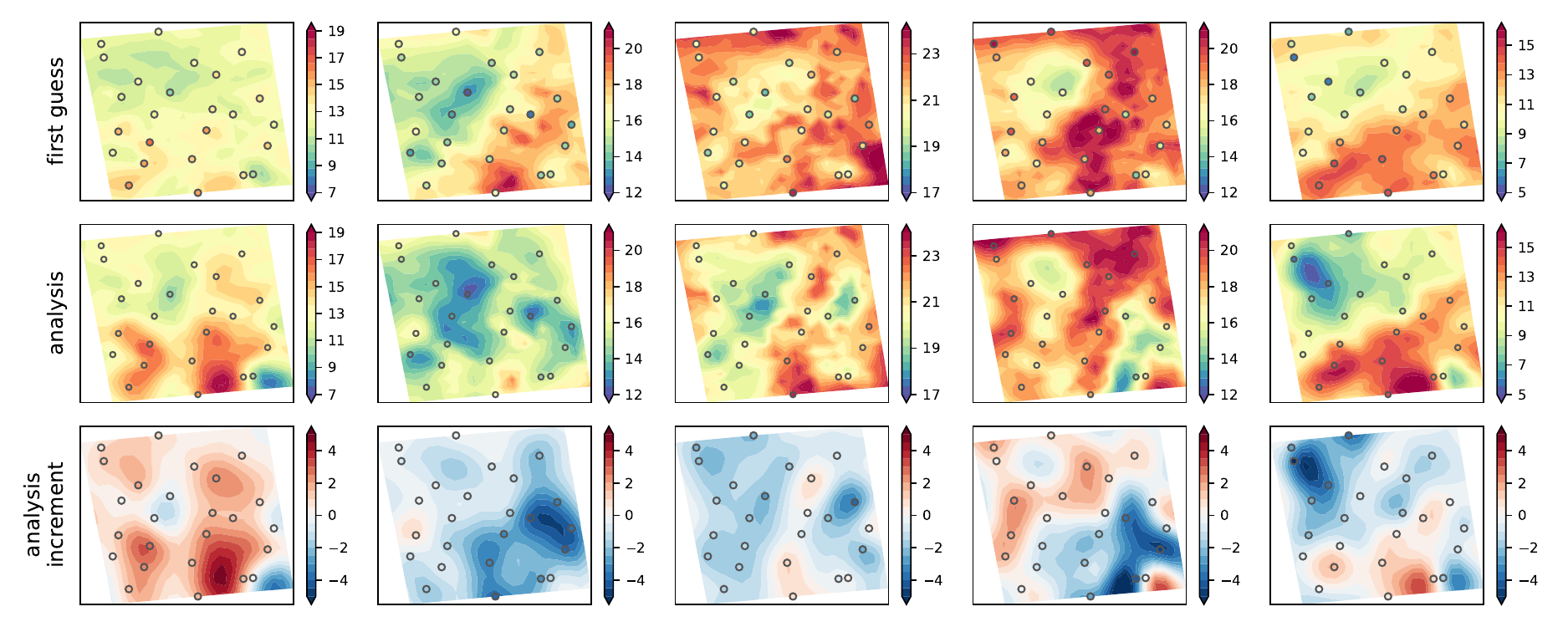}
	\caption{Five examples from the validation data of the real case experiment. The contours show the first guess field in the top row and the AI-Var analysis in the center row. The circles denote the respective T2M observations. The bottom row depicts the contours of the analysis increments and the deviation of the observations from the first guess in the circles.}
 	\label{fig:t2m_examples}
\vspace{0mm}
\end{figure}
%--------------------------------------------------------------------------------------

\subsection{Real test case}
\label{sec:results_real}

Using the experiment setup described in section \ref{sec:realsetup}, we apply our AI-Var to a real world example for data assimilation.

As usual, the loss is decreasing with increasing epochs as shown in Figure \ref{fig:t2m_loss}. Here, training and validation loss move in parallel, i.e., an improvement in the model as indicated by a lower training loss also leads to a decrease of the validation loss. However, with the training loss still trending downward induced by the steep reduction of the observation loss term, the model is not yet fully trained and may further improve as there is no evidence of overfitting yet, i.e., the validation loss keeps decreasing, too. Due to the complexity, a larger number of training epochs may be reasonable. However, as the following results indicate, the model reaches a sufficient level of quality for your proof-of-concept approach.

Figure \ref{fig:t2m_examples} shows the application of the trained model to the validation data set. The contour plots depict the first guess fields from the reanalysis (top row) and the analysis from our AI-Var (center row). The circles denote the observation locations with the colors indicating the observed value in the same color range as the contour plots. The bottom row depicts the filled contours of the respective analysis increments with the circles showing the deviation of the observations from the first guess.

While some observations agree with the first guess fields, there is often a mismatch between the observed and estimated T2M in the first guess. From the center row, we find that AI-Var is able to locally correct for these deviations such that the observation match the analysis field very well. This can also be deduced from the bottom row where we find that for almost all observations, the analysis increment matches the error of the first guess to the observed values.

%--------------------------------------------------------------------------------------
% FIGURE - T2M DIURNAL CYCLE
%--------------------------------------------------------------------------------------
\begin{figure}[t]
\vspace{-5mm}
  \centering
    \includegraphics[width=0.49\textwidth]{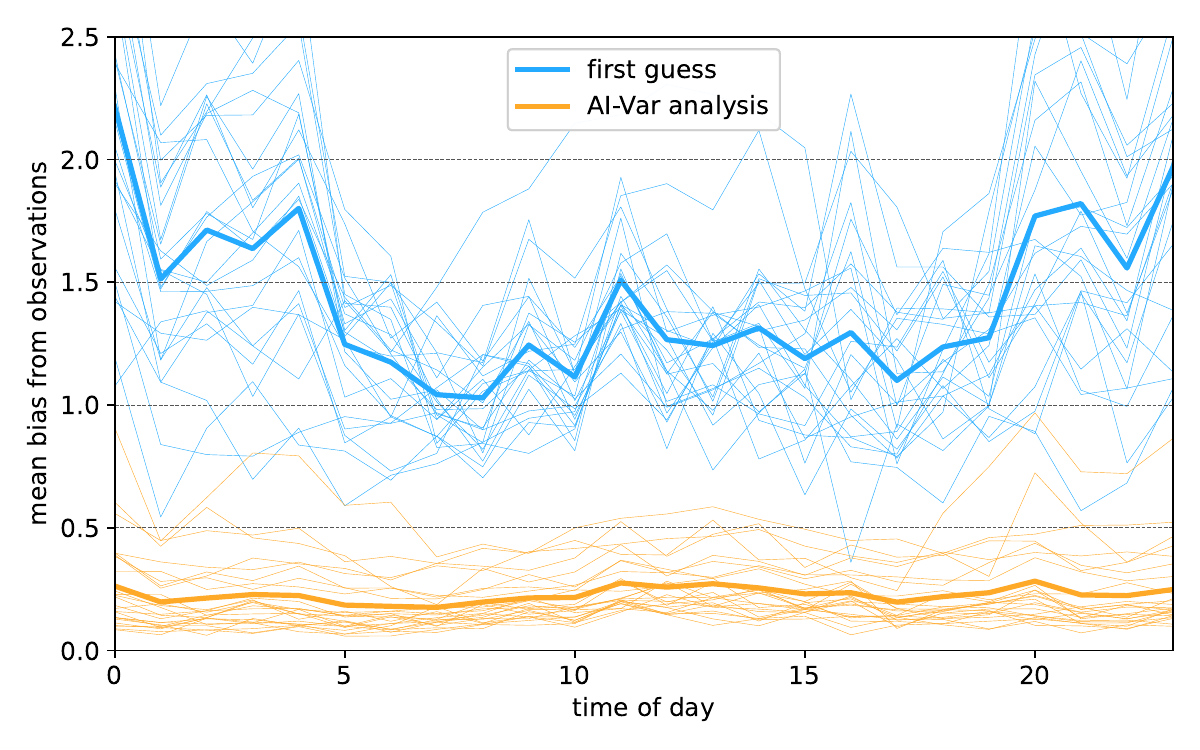}
    \includegraphics[width=0.49\textwidth]{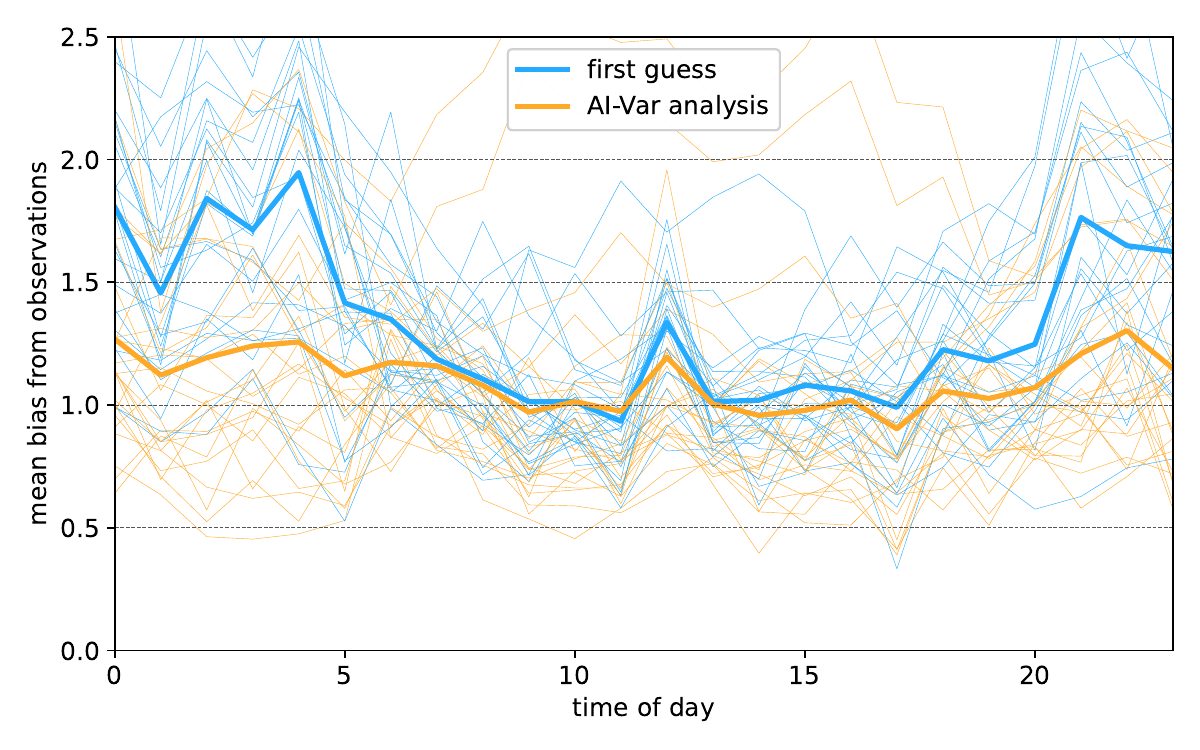}
	\caption{Diurnal cycle of first guess (blue) and AI-Var analysis (yellow) deviations from the observations in the validation data set. Thin lines indicate the individual $25$ observation sites whereas the thick line denotes the mean over all stations. The left plot shows the experiment using all observations, the right plot the cross-validation experiment.}
 	\label{fig:t2m_diurnal}
\vspace{-5mm}
\end{figure}
%--------------------------------------------------------------------------------------

Another evaluation is given in the left plot in Figure \ref{fig:t2m_diurnal} which depicts the diurnal cycle of the mean absolute deviations of the observations from the first guess (blue) and AI-Var analysis (yellow). The single observations are denoted as thin lines, whereas the thick lines are the average over all observations. The upper plot shows the result for the validation period of our real case experiment. We find that there is a clear diurnal cycle of first guess errors with respect to the observations  with larger deviations becoming larger during the night time. This is due to a nightly warm bias in the COSMO-REA6 reanalysis which is well known. However, with our AI-Var, we are not only able to significantly reduce the error but also eliminate the diurnal cycle of the T2M errors from the first guess field ($80\%$-$90\%$ reduction).

In our cross-validation experiment for the real case, we try to evaluate the merit of our AI-Var system using independent observations. The evaluation of the five training runs withholding five different observations from the full observational data set in each run, are combined to confirm the validity of our approach. We find a reduction of root mean square error from first guess to analysis in $20$ of the $25$ observations with an average reduction of $26\%$ in these $20$ observations and an increase of about $16\%$ in the remaining $5$ observations. It is important to note that as a proof-of-concept, we only employed a Gaussian B-matrix which might not be an optimal choice for such heterogeneous fields as T2M. Despite this strong constraint, the AI-Var is still able to enhance the T2M representation in the cross-validation experiment.

We find further evidence of this in the right hand side plot in Figure \ref{fig:t2m_diurnal} which shows the results from the cross-validation experiments for the diurnal cycle of the deviations form the observations. Here, the differences are calculated from the independent observations from the five trained models, respectively. While for some observations the errors even increase in the analysis, on average we still find a reduction of the deviations through data assimilation and an attenuation of the nightly warm bias. It is important to note that as a proof-of-concept, we only employed a Gaussian B-matrix which might not be an optimal choice for such heterogeneous fields as T2M. Despite this strong constraint, the AI-Var is still able to enhance the T2M representation in the cross-validation experiment.

\section{Conclusions}
\label{sec:conclusions}

In this paper, we have presented a novel AI-based data assimilation approach that has the potential to replace the classical data assimilation scheme by training a neural network to perform the data assimilation task itself. This proof-of-concept study has shown promising results across both idealized and real-world scenarios. 

Our investigations in one-dimensional (1D) and two-dimensional (2D) idealized cases indicate that the AI-Var approach is - even with a zero first guess - capable of effectively reconstructing the original state from observations. In particular, we have noticed that while the reconstruction can be nearly perfect for static observation setups, the performance decreases only modestly for the more challenging dynamic observation locations case. This indicates that combining a sufficient number of observations with a large training data set can reasonably compensate for the added complexity introduced by dynamic observation locations.

Our real-world test case, which involved the assimilation of 2-meter temperature (T2M) observations from synoptic observation sites, has demonstrated that our AI-Var system can indeed improve estimation accuracy. Specifically, it can correct for biases in the first guess fields, such as the nocturnal warm bias in the COSMO-REA6 reanalysis, and reduce errors significantly. These improvements were to a lesser extent also evident in the cross-validation experiment with independent observations, though the Gaussian B-matrix employed may not be the optimal choice for heterogeneous fields like T2M.

As with any novel approach, there still remains a large potential for improvement with further research and development necessary before AI-based data assimilation systems can be fully operational. Future work may focus on:
First, {\em enhanced neural architectures} are needed to handle the increased complexity of high-dimensional data assimilation problems.
Second, {\em advanced training strategies} including iterative training approaches and meta-learning will be needed to further reduce computational costs and improve the quality of solutions.
Third, {\em realistic 3D applications} will be an important next step extending the approach to three-dimensional as well as multi-variate data sets to better emulate real-world meteorological scenarios and increase the realism of the results.
Fourth, {\em optimized covariance models} have been important to improve quality in traditional algorithms. Investigating alternative B-matrix representations beyond the Gaussian kernel such as climatological and ensemble-based approaches to better capture the variability and heterogeneity of different meteorological variables will be an important development step. We note that from a conceptional viewpoint the use of an ensemble based B-Matrix in the loss functional (\ref{eqn:3dvar}) as in the EnVar \citep[e.g., ][]{Buehner2013} is straightforward. This needs, however, the generation of an appropriate ensemble and, thus, if one aims to have a fully data-driven analysis, further progress in AI based ensemble generation. 
Fifth, it is important to explore the {\em operational feasibility} of AI-based data assimilation, integrating this approach into existing NWP systems, possibly through hybrid data assimilation strategies that blend traditional and AI-based methods.

In conclusion, our AI-based data assimilation method shows a clear path forward for leveraging machine learning to handle traditionally complex and computationally expensive tasks. By fully integrating neural networks into the assimilation process, we open up opportunities for faster, more efficient, and eventually more accurate weather prediction systems—paving the way for fully data-driven NWP systems in the future.

\section*{Disclaimer}
This Work has not yet been peer-reviewed and is provided by the contributing Author(s) as a means to ensure timely dissemination of scholarly and technical Work on a noncommercial basis. Copyright and all rights therein are maintained by the Author(s) or by other copyright owners. It is understood that all persons copying this information will adhere to the terms and constraints invoked by each Author's copyright. This Work may not be reposted without explicit permission of the copyright owner.

\appendix

\section{AI-Var Pseudo Code}
\label{sec:appendix_psuedo_code}

\begin{enumerate}[leftmargin=2cm, label=\arabic*.]

    \item \textbf{Set Input Data:}
    \begin{lstlisting}
z := nt x ng array containing the grid input data on the grid
o := nt x no array containing the observations input data
i := nt x no array containing the locations of the observations on the grid
    \end{lstlisting}

    \item \textbf{Define Input Fields and Vectors for Training:}
    \begin{lstlisting}
sort_indices = sort i in ascending order
input_field  = scale and convert z to tensor
input_vector = stack sorted o and normalized i into tensor
b_cov        = calculate covariance matrix with Gaussian kernel width sigma
regularize b_cov
\end{lstlisting}

    \item \textbf{Initialize Training Data and DataLoader:}
    \begin{lstlisting}
train_set    = create training dataset with input_field and input_vector
train_loader = create DataLoader for training with batch_size and shuffling based on train_set
    \end{lstlisting}
    
    \item \textbf{Define Data Assimilation Neural Network and Loss Function:}
    \begin{lstlisting}
class DataAssimilationNN:
    initialize with dimensions ng, n_obs, obs_error
    define three fully connected layers
    forward pass:
        concatenate input first guess, observations and their locations
        apply relu activation after each fully connected layer
        reshape and return the output
        
class DataAssimilationLoss:
    initialize with ng, b_matrix, obs_err
    forward pass:
        calculate background mismatch using Mahalanobis distance
        sample observations from model output
        calculate observation mismatch
        combine losses and return
    \end{lstlisting}

    \item \textbf{Prepare Training:}
    \begin{lstlisting}
model      = initialize DataAssimilationNN with ng, n_obs
criterion  = initialize DataAssimilationLoss with regularized_b_cov, ng, obs_err
optimizer  = initialize Adam optimizer with learning rate 0.001
    \end{lstlisting}

    \item \textbf{Training Loop:}
    \begin{lstlisting}
for epoch in range(num_epochs):
    for batch in train_loader:
        clear optimizer gradients
        output  = model(batch)
        loss    = criterion(output, batch)
        backpropagate loss
        update model parameters with optimizer
        store training loss
    \end{lstlisting}

\end{enumerate}

\bibliographystyle{abbrvnat}
\bibliography{references}

\end{document}